\documentclass[aps,prb,twocolumn,showpacs,preprintnumbers,amsmath,amssymb]{revtex4}
\usepackage{graphicx}% Include figure files
\usepackage{dcolumn}% Align table columns on decimal point
\usepackage{bm}% bold math

\begin{document}

\preprint{APS/123-QED}

\title{Nonionic contributions to the electric-field gradient at $^{181}$Ta and $^{111}$Cd impurity sites
in R$_2$O$_3$ (R= Sc, In, Lu, Yb, Tm, Er, Y, Ho, Dy, Gd, Eu, Sm) bixbyites }% Force line breaks with \\

\author{Leonardo A. Errico}%\email[Correspondace should be addressed to:
%]{renteria@fisica.unlp.edu.ar; errico@fisica.unlp.edu.ar}
\author{Mario Renter\'{\i}a}%\altaffiliation[Email Address: ] {renteria@venus.fisica.unlp.edu.ar}
\email[Correspondace should be addressed to: ]
{renteria@fisica.unlp.edu.ar; errico@fisica.unlp.edu.ar}
%\altaffiliation[Also at ]{Instituto de F\'{\i}sica La Plata
%(CONICET).}
\author{An\'{\i}bal G. Bibiloni}
%\altaffiliation[Also at ]{Instituto de F\'{\i}sica La Plata
%(CONICET).}

\affiliation{Departamento de F\'{\i}sica, Facultad de Ciencias
Exactas, Universidad Nacional de La Plata, C.C. 67, 1900 La Plata,
Argentina.}
%\\ This line break forced

\author{Kristian Freitag}
\affiliation{Helmholtz - Institut f\"{u}r Strahlen- und Kernphysik
(ISKP), Universit\"at Bonn, Nussallee 14-16, 53115 Bonn, Germany}

\date{\today}% It is always \today, today,

\begin{abstract}
The time-differential perturbed-angular-correlation (TDPAC)
technique was applied to the study of the internal electric-field
gradient (EFG) in Eu- and Ho-sesquioxides in their cubic bixbyite
phases.
 The results, as well
as previous characterizations of the EFG at $^{181}$Ta sites in
oxides with the bixbyite structure, were compared to those
obtained in experiments using $^{111}$Cd as probe, and to
point-charge model and {\it ab initio} results calculations for
the EFG tensor at impurity sites in binary oxides. These studies
provide quantitative information about electronic processes and
the structural relaxations induced by the presence of impurity
probes in the host lattices, and confirm the existence of nonionic
contributions to the EFG in these systems. Our FP-LAPW
calculations show that this nonionic contribution to the EFG is
the dominating one, and that it is originated in the population of
{\it p} states (5{\it p} in the case of Cd, 6{\it p} for Ta).

\end{abstract}


\maketitle

\section{\label{sec:1}INTRODUCTION}

During the last three decades, time-differential $\gamma-\gamma$
perturbed-angular-correlation (TDPAC) spectroscopy has been
increasingly applied to condensed matter problems through the
precise characterization of the electric-field-gradient (EFG)
tensor at diluted (ppm) radioactive probe atoms, adequately
introduced in substitutional host lattice sites (see, e.g.,
Refs.~\onlinecite{KauVianden,Lerf,Abeti} and references therein).
The characteristics of TDPAC allow carrying out the experiments
under a wide range of suitable external conditions such as
variable temperatures, pressures, atmospheres, etc. The
single-atom counting of this technique, in combination with its
highly localized sensitivity (due to the r$^{-3}$ dependence of
the electric-quadrupole interaction), allows a detailed
investigation of both structural and electronic properties of the
systems, providing information about crystal
structures,\cite{Lupascu94a} crystal chemistry,\cite{Catchen94}
defects, \cite{Lany2000,Evenson2000} nanoscopic characterizations
of surfaces, interfaces,\cite{Schatz86} and highly dispersed
species\cite{Ramallo2003,Pasque83}, among other  properties in
solid state
physics\cite{Errico99,Bibi84,Habe96,Achtziger93,Vianden1988,Renteria2000,
Forker2003,Rots2001}, chemistry and biology (see, e.g.,
Ref.~\onlinecite{Lerf1997}, and references therein).

The very well suited ($^{111}$In $\rightarrow$)$^{111}$Cd isotope
is the most frequently used tracer in TDPAC experiments and has
been largely applied to study semiconductor physics. In
particular, after the initial work of Pasquevich {\it et
al.}\cite{Pasque81} on the internal oxidation of diluted indium
impurities in Ag, a large amount of experimental work has focused
on the EFG characterization at $^{111}$Cd impurity sites in
semiconductor and insulating binary oxides.\cite{Wiarda92} All the
information that the EFG tensor can provide about the system under
study could be obtained by confrontation of the experiment with an
accurate prediction of the EFG, such as those obtained with {\it
ab initio} calculations. In the absence of such predictions at
impurity sites, several attempts to correlate experimental results
and semiempirical calculations have been made from the very
beginning,\cite{Bolse88,Kesten90} in order to describe the
different contributions to the EFG at impurity
sites.\cite{RenteThesis}   In 1992, a well-defined empirical
correlation between local (Cd valence states) and ionic
contributions to the EFG was presented for $^{111}$Cd in binary
oxides.\cite{Renteria92} In that analysis, the local component of
the EFG in binary oxides was extracted from all
quadrupole-coupling constants measured at that time in TDPAC
experiments with the $^{111}$Cd probe. The resulting systematics
revealed a linear dependence between the local and ionic
contributions to the EFG over a wide range of ionic EFG values. In
1994, Weht {\it et al.}\cite{Weht94} confirmed all the features of
the correlation through an independent cluster calculation of the
valence EFG contribution using the Extended H\"ueckel Method.
Recently, Errico {\it et al.}\cite{PRL,PRB} reported the first
{\it ab initio} full-potential linearized-augmented-plane-wave
(FP-LAPW) calculations of the EFG at an impurity site (Cd) in an
oxide (TiO$_2$), in excellent agreement with the experiments,
supporting the existence of a dominant valence contribution to the
EFG in this system, which can be identified with the local
contribution of Ref.~\onlinecite{Renteria92}.

To study the influence of the electronic configuration of the
impurity probe atom itself on the EFG it is essential to perform
TDPAC measurements with different probes in isomorphus crystal
structures. Examples of this kind of study are the work of Adams
and Catchen, \cite{Catchen94} and Shitu et al.\cite{Shitu98} In
TDPAC experiments, the second most commonly used radioactive probe
is $^{181}$Hf, which decays by $\beta^{-}$ to the $^{181}$Ta
isotope. Therefore, in 1990 we started a comparative study of the
EFG at $^{181}$Ta and $^{111}$Cd sites in binary oxides with the
aim of investigating the EFG dependence on the electronic
configuration of the probe atom and its coordination ``chemistry''
with its nearest neighbors.\cite{Moreno91} Among the binary
oxides, the sesquioxides that crystallize in the cubic bixbyite
structure constitute a group widely studied with the $^{111}$Cd
probe. \cite{Errico99,Bibi84,Bartos91}$^-$\cite{Carbonari99} This
group presents interesting properties since it includes a large
series of rare-earth as well as closed-shell metallic cations, a
wide range of lattice constants (from 0.94 to 1.09 nm), and two
nonequivalent cation sites with quite different symmetry,
cation-oxygen bond lengths, and crystallographic abundance. These
characteristics make this group ideally suited to study in detail
the EFG dependence on the coordination geometry of the probe atoms
and on their electronic configuration. With this purpose, in 1994
we started a systematic TDPAC study of this group using $^{181}$Ta
as
probe.\cite{Pasque94,Renteria97,Renteria98,Renteria99,Errico2001,Germantulio}
From this systematic study and using the same hypothesis as in
Ref.~\onlinecite{Renteria92}, we found an empirical correlation
between the local and ionic contributions to the EFG similar to
those found for $^{111}$Cd, but with different
slope.\cite{Renteria99}

In this work we report TDPAC experiments using $^{181}$Ta as probe
in Eu$_2$O$_3$ and Ho$_2$O$_3$ in order to complete the systematic
study of the EFG at $^{181}$Ta atoms located at defect-free cation
sites in the bixbyite structure. This systematics is compared to
those previously established for the $^{111}$Cd probe and to
recent and new {\it ab initio} band-structure calculations of the
EFG at impurity sites in a small set of binary oxides. This
comparison provides quantitative information about electronic
processes and the structural relaxations induced by the presence
of impurity probes in the host lattices, information that cannot
be obtained (or is crudely estimated) by simple models such as the
point-charge model, the use of antishielding factors, and
arbitrary suppositions. The {\it ab initio} calculations also
predict the existence of a dominating nonionic (or local)
contribution to the EFG originated from Cd-5$p$ and Ta-6$p$
states, in good agreement with the semiempirical model proposed in
references ~\onlinecite{Renteria92} and ~\onlinecite{Renteria99}.
These new theoretical results, in combination with the
experimental ones, enable us to discuss the validity of the widely
used ionic model and to validate or discard the hypothesis used in
the construction of the semiempirical model mentioned above and
its predictions.

\section{SAMPLE PREPARATION AND TDPAC MEASUREMENTS}

Under suitable conditions, Fe, Mn, Sc, In, Tl, Y, and all the
rare-earth elements form a sesquioxide. Polymorphism is common
among the rare-earth oxides and below about 2300 K three
polymorphous crystallographic structures have been
found:\cite{Eyring79} the hexagonal A-, the monoclinic B-, and the
cubic C-form (bixbyite). In the cubic structure the cations form a
nearly cubic face-centered lattice (space group Ia3) in which six
out of the eight tetrahedral sites are occupied by oxygen atoms.
The elementary cell of the oxide lattice consists of eight such
cubes, containing 32 cations and 48 Oxygen ions. Two nonequivalent
cationic sites, called $C$ and $D$, both O$_6$-coordinated,
characterize the structure. Their relative abundance in the
lattice is ($f_C/f_D) = 3$. Site $D$ is axially symmetric and can
be locally described (see Fig. 1) as a cation surrounded by six
oxygen atoms localized at the corners of a distorted cube, leaving
two corners of a diagonal free (D$_{3d}$ point-group symmetry). At
site $C$ the cube is more distorted (C$_2$-symmetry) and the six
oxygen atoms leave two corners on a face diagonal free (see Fig.
1).

\begin{figure}
\includegraphics*[bb= 61mm 224mm 125mm 263mm, viewport=-1.2cm 0cm 8.5cm 4.8cm, scale=1]{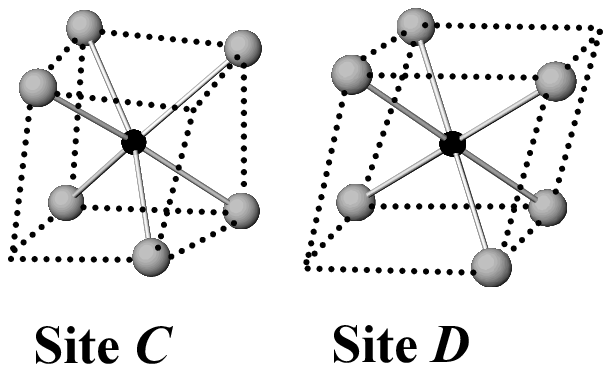}
\caption{\label{fig:1} Nearest-neighbor (NN) oxygen distribution
around each cationic site (black atom) in the bixbyite structure.}
\end{figure}

In order to prepare the TDPAC samples, commercially obtained
high-purity Ho$_2$O$_3$ and Eu$_2$O$_3$ powders (Aldrich Products,
99.9\% and 99.95\% metallic purity, respectively) were treated in
air for 24 h at 1023 K in order to achieve the crystalline C-phase
and then pressed as circular pellets under 20 kN/m$^2$ pressure.
Afterward, the Ho$_2$O$_3$ pellet was sintered in air for 2.5 h at
1273 K whereas Eu$_2$O$_3$ was sintered under similar conditions
but at 1123 K for 1.5 h because in this oxide the C-form presents
a phase transition to the monoclinic B-phase at around 1353
K.\cite{Eyring79} In both cases, powder XRD analyses of the
samples performed before and after the sintering showed that only
the C-phase was present and also revealed an increase in the
crystallinity of the samples. The ion accelerator of the ISKP,
Bonn, was then used to implant $^{181}$Hf$^+$ ions into the
samples with the following energies and doses: Ho$_2$O$_3$, 160
keV, 5x10$^{13}$ ions/cm$^2$ and Eu$_2$O$_3$, 155 keV, 5x10$^{12}$
ions/cm$^2$. The as-implanted samples were subjected to TDPAC
measurements in air at room temperature (RT = 300 K) and at
atmospheric pressure. After these measurements the Ho$_2$O$_3$ and
Eu$_2$O$_3$ samples underwent an annealing treatment in air at
1323 K and 1123 K for 2 h, respectively, since in other rare-earth
sesquioxides annealing above 1073 K for a couple of hours has
demonstrated to be sufficient to restore the host crystallinity
and to locate the impurities at defect-free cationic sites. In the
case of Ho$_2$O$_3$ the TDPAC measurements were then carried out
in air in the temperature range RT-1373 K in 100 K steps. In the
case of Eu$_2$O$_3$ the measurements were performed between RT and
1273 K (in 100 K steps) and at 77 K.

\begin{figure}[t]
\includegraphics*[bb= 42mm 165mm 120mm 263mm, viewport=0.2cm 0cm 8.5cm 11cm, scale=0.95]{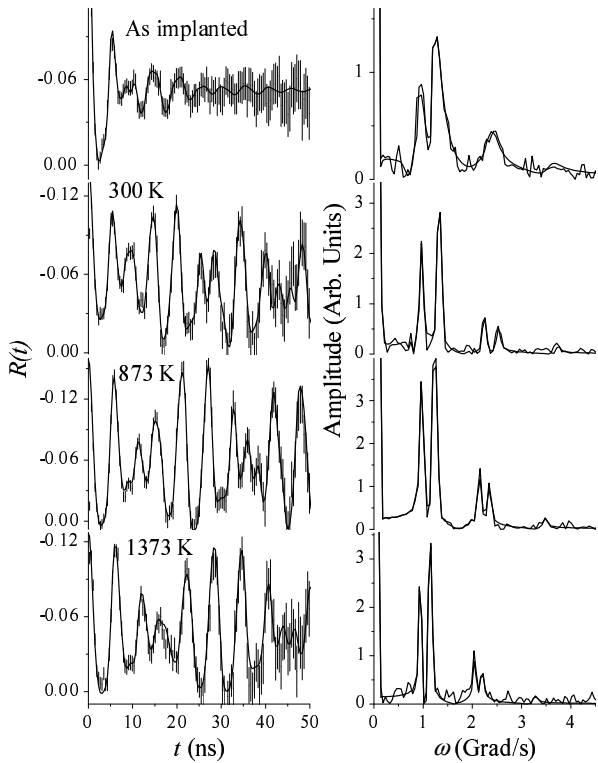}
\caption{\label{fig:2}  $R(t)$ spectra (left) and their
corresponding Fourier transforms (right) of
($^{181}$Hf$\rightarrow$)$^{181}$Ta in Ho$_2$O$_3$ measured at RT
in the as-implanted stage and at the indicated temperature after
thermal annealing in air at 1323 K for 2 h.}
\end{figure}

The TDPAC technique is based on the determination of the influence
of extranuclear fields on the correlation between the emission
directions of two successive radiations emitted during a
nuclear-decay cascade. A complete description of this technique
can be found in the literature (see, e.g.,
Refs.~\onlinecite{Lerf1997} and~\onlinecite{Frauen}). In order to
perform the experiments presented in this work we made use of the
well-known 133-482 keV $\gamma-\gamma$ cascade in $^{181}$Ta,
produced after the $\beta^-$ nuclear decay of the $^{181}$Hf
isotope. The TDPAC experiments were made using four BaF$_2$
detectors in a coplanar 90$^{\circ}$ arrangement and a fast-fast
logic coincidence system. The time and energy resolutions of the
spectrometer for $^{181}$Ta $\gamma$-rays are 0.65 ns (full width
at half-maximum of the prompt peak) and 9.5\% for $\gamma$-rays of
662 keV ($^{137}$Cs), respectively. The experimental perturbation
functions, $R(t)$, were derived from eight concurrently measured
coincidence spectra, four taken between detectors positioned with
180$^{\circ}$ symmetry and four of eight possible with
90$^{\circ}$ symmetry.\cite{Renteria97} To analyze the measured
perturbation functions, a multiple-site model for
nuclear-electric-quadrupole interactions for polycrystalline
samples and spin I=5/2 intermediate level of $^{181}$Ta was used

\begin{eqnarray}
R(t) = A_{22}^{exp}G_{22}(t)= \nonumber\\
= A_{22}^{exp}\sum_{i} f_i { S_{20,i} + \sum_{n=1}^{3}[S_{2n,i}
\cos(\omega_{n,i}t)e^{-\delta_i \omega_{n,i}t } ] },\label{G22}
\end{eqnarray}
where $f_i$ is the relative fraction of nuclei that experiences a
given perturbation and $A_{22}^{exp}$ is the effective anisotropy
of the $\gamma-\gamma$ cascade. The $\omega_n$ interaction
frequencies are related to the quadrupole frequency
$\omega_Q=eQV_{ZZ}/40\hbar$ by $\omega_{n}=g_{n}\omega_Q$. The
$g_n$ and $S_{2n}$ coefficients are known
functions\cite{Mendoza77} of the asymmetry parameter
 $\eta=(V_{XX}-V_{YY})/V_{ZZ}$, where $V_{ii}$ are the principal
  components of the EFG tensor that are
arbitrarily labeled according to $|V_{XX}|<|V_{YY}|< |V_{ZZ}|$.
The exponential functions in Eq.~\ref{G22} account for a
Lorentzian frequency distribution of relative width $\delta$
around $\omega_n$.

\begin{figure}
\includegraphics*[bb= 43mm 165mm 130mm 272mm, viewport=0cm 0cm 8.5cm 11cm, scale=0.95]{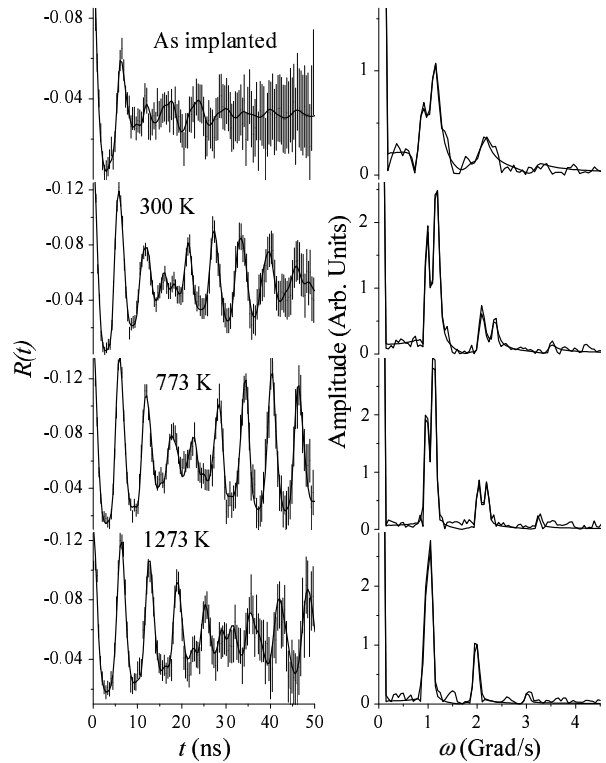}
\caption{\label{fig:3} $R(t)$ spectra (left) and their
corresponding Fourier transforms (right) of
($^{181}$Hf$\rightarrow$)$^{181}$Ta in Eu$_2$O$_3$ measured at RT
in the as-implanted stage and at the indicated temperature after
thermal annealing in air at 1123 K for 2 h.}
\end{figure}

\section{\label{sec:3}$^{181}$Ta TDPAC RESULTS IN Ho$_2$O$_3$ and Eu$_2$O$_3$}

\begin{figure}
\includegraphics*[width=3.15in]{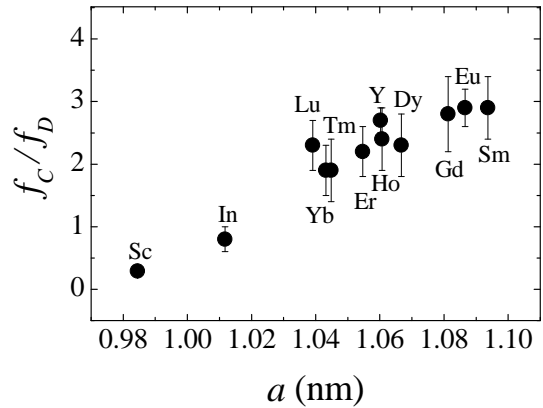}
\caption{\label{fig:4} Experimental ratio $f_C/f_D$ as a function
of the lattice parameter $a$ for $^{181}$Ta in bixbyites.}
\end{figure}

 Figures ~\ref{fig:2} and \ref{fig:3} show the $R(t)$ spectra and their
  corresponding Fourier transforms,
taken on both as-implanted samples at RT in air, and the selected
spectra taken after annealing of the samples. Solid lines in the
$R(t)$ spectra are the best least-squares fits of Eq.~\ref{G22} to
the experimental data. Solid lines in the Fourier spectra come
from the Fourier transform of the $R(t)$ fits. As can be seen from
both figures, the as-implanted $R(t)$ spectra are rather damped,
as usually occurs in the as-implanted stage of
$^{181}$Hf-implanted binary
oxides,\cite{Renteria97,Pasque94,Renteria98,Errico2001} showing
that the radiation damage produces a not negligible host disorder
around the probe atoms. After the annealing treatments, the
radiation damage was removed in both samples. In effect, only  two
very well-defined ($\delta<1.5\%$) hyperfine interactions (labeled
Hfi $C$ and Hfi $D$) were found, with the asymmetry parameter of
Hfi $D$ equal to zero and a high $\eta$ value for Hfi $C$ (see
Table~\ref{table1}) as predicted by the coordination symmetries of
sites $D$ and $C$, respectively. In addition, the frequency of the
axially symmetric interaction is twice as large as the asymmetric
one, as found previously with the $^{181}$Ta probe in Yb-, Y-,
Dy-, Sm-, Lu-, Gd-, Er- and
Tm-sesquioxides.\cite{Pasque94}$^-$\cite{Germantulio} The
asymmetry parameters and distributions of Hfi $C$ and Hfi $D$ for
both oxides are constant with temperature, while the quadrupole
frequencies show  a reversible continuous decrease with increasing
temperature in the measured temperature range. The monotonic trend
of the hyperfine parameters of Hfi $C$ and Hfi $D$ is a typical
behavior followed by hyperfine probes in substitutional cation
sites  of a single-phase crystalline structure. The total fraction
of both interactions amounts to 100\% of the probes in the whole
temperature range of measurement for Eu$_2$O$_3$. This statement
also applies to Ho$_2$O$_3$, with the exception that two minor
interactions amount to less than 20\% of the probes in the
temperature range RT-472 K. The relative population of the two
sites should be ($f_C/f_D) = 3$ if the cationic sites were
occupied according to their natural abundance in the crystalline
structure. In the present experiments, ($f_C/f_D) = 3$ was found
for Eu$_2$O$_3$. In the case of Ho$_2$O$_3$, the experimental
ratio $f_C/f_D$ was smaller than this value. This departure was
already observed for ($^{181}$Hf$\rightarrow$)$^{181}$Ta in other
bixbyites and was explained in terms of the relative ionic size of
the probe and the cationic ``space'' of the
host.\cite{Renteria97,Renteria98} The present results for Ho- and
Eu-sesquioxides are in excellent agreement with the occupancy
trend previously observed (see Fig.~\ref{fig:4}). It is then clear
that Hfi $C$ and Hfi $D$ can be undoubtedly assigned to probes
located at the crystalline nonequivalent cationic sites $C$ and
$D$.

We want to mention that the $f_C/f_D$ ratio for Ho$_2$O$_3$ and
Eu$_2$O$_3$ obtained with $^{181}$Ta, as well as those for the
rest of the bixbyite series, is independent of the sample
temperature (in the range RT-1300 K). This may be evidence of no
further ionic transport after the annealing process.

The two minor frequencies observed for Ho$_2$O$_3$ exist in a
short temperature range and disappear above 573 K in a reversible
way. Due to the temperature range in which these minor
interactions exist we have investigated the possibility  of
desorption-absorption of water in the samples. However, a
Differential-Thermal Analysis (DTA) carried out for the
Ho$_2$O$_3$ samples did not confirm this hypothesis.

%\squeezetable
\begin{table*}
\caption{\label{table1} Results of least-squares fits of
Eq.~\ref{G22} to the $R(t)$ spectra displayed in Figs.~\ref{fig:2}
and \ref{fig:3}, taken at RT after annealing. $f_i$ and $\delta_i$
are expressed in \% and $\omega_{Q i}$ in Mrad/s. When no errors
are quoted it means that the parameter was kept fixed in order to
evaluate errors. In the case of Ho$_2$O$_3$ the fractions $f_C +
f_D$ are normalized to 100\% (see text).}
\begin{ruledtabular}
\begin{tabular}{llllllllllll}
  & \multicolumn{4}{c}{Hfi $C$} & & &\multicolumn{4}{c}{Hfi $D$}\\
     & \multicolumn{1}{c}{$f$} & \multicolumn{1}{c}{$\omega_Q$} &
  \multicolumn{1}{c}{$\eta$} & \multicolumn{1}{c}{$\delta$} &
  & \multicolumn{1}{c}{$f$} & \multicolumn{1}{c}{$\omega_{Q}$} &
   \multicolumn{1}{c}{$\eta$} & \multicolumn{1}{c}{$\delta$} \\ \hline

 Ho$_2$O$_3$& $70(4)$ & $114.3(3)$ & $0.612(4)$ & $0.0(2)$&    & $30(5)$
& $207.0(7)$ & $0$ & $0.3(4)$\\ \hline

  Eu$_2$O$_3$& $74(8)$ & $101.2(3)$ & $0.798(4)$ &
$1.6(4)$&  & $26(3)$ & $193.8(9)$ & $0$ & $1(1)$
\end{tabular}
\end{ruledtabular}
\end{table*}

\section{\label{sec:4}DISCUSSION}

\subsection{\label{sec:4.A}Systematics of the EFG at  $^{111}$Cd and $^{181}$Ta
impurity sites in bixbyites}

\begin{figure}
\includegraphics*[bb= 10mm 165mm 200mm 273mm, viewport=0.8cm 0.7cm 10.2cm 11cm, scale=0.9]{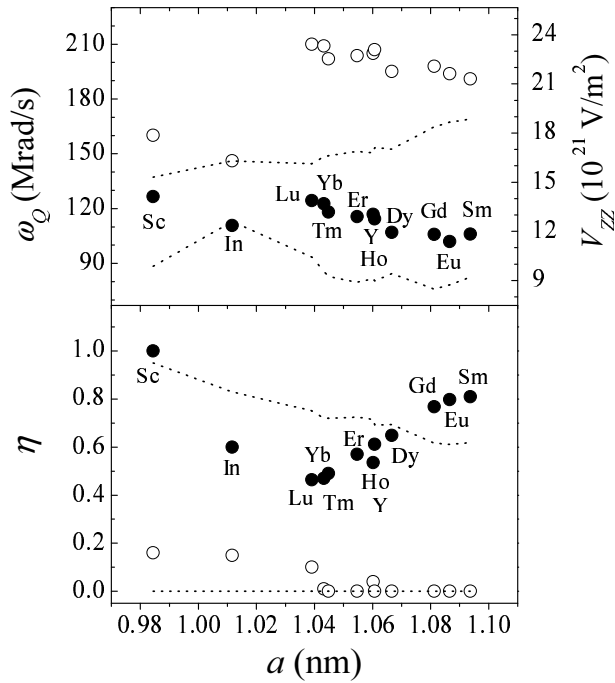}
\caption{\label{fig:5} Experimental quadrupole frequency
$\omega_Q$ and asymmetry parameter $\eta$ for substitutional
$^{181}$Ta impurities at sites $C$ (black) and $D$ (hollow) in
bixbyites plotted as a function of the lattice parameter $a$ at
RT. To obtain $V_{zz}$ from $\omega_Q$ we used $Q_{Ta} = (+)
2.36(5)~b$.\cite{Butz83} The errors are smaller than the symbols.
Dotted lines are the PCM predictions. The $\omega_Q$ predictions
are normalized to the In$_2$O$_3$ values.The lattice parameters
and the atomic coordinates used in the PCM calculations for each
oxide have been taken from neutron and x-ray diffraction (XRD)
determinations quoted in the literature.\cite{DRXN}}
\end{figure}

    In 1994, we started a TDPAC study of the EFG dependence on the probe atom in bixbyites, measuring the EFG at
$^{181}$Ta impurity sites in Yb-, Y-, and
Dy-sesquioxides,\cite{Pasque94} In$_2$O$_3$,\cite{Renteria97}
Sc$_2$O$_3$ and Sm$_2$O$_3$,\cite{Renteria98,Renteria99}
Lu$_2$O$_3$,\cite{Renteria99}  Er$_2$O$_3$
  and Gd$_2$O$_3$,\cite{Errico2001} and Tm$_2$O$_3$.\cite{Germantulio}
   Including the new data presented in this paper,
  i.e., Eu$_2$O$_3$ and Ho$_2$O$_3$, we can now compare a set of twelve
oxides, all having the same crystalline structure, but differing
in their lattice constant by 10\%.
    In Fig.~\ref{fig:5} we present the values of $\omega_Q$ and $\eta$ for both sites
 as a function of the lattice parameter $a$. All the experimental
 data in this figure (and also in Fig 4) correspond to TDPAC
 experiments in which the $^{181}$Hf isotope was introduced into the
 samples by means of ion implantation, with the exception of Tm$_2$O$_3$. In this
 case the activity was introduced by means of a solid-state reaction
 assisted by ball milling between powders of Tm-sesquioxide and neutron-activated HfO$_2$.
 In spite of the limited accuracy of the derived hyperfine parameters in this experiment,
 due to a remaining 40\% fraction of probes located in a highly distorted
 HfO$_2$ phase, the hyperfine interaction parameters corresponding to Tm$_2$O$_3$ appear
 to be reliable considering their excellent agreement with the rest of the
 systematics. As can be seen in Fig. 5, there is a jump in the $^{181}$Ta
systematics. In effect, $\omega_{QC}$ and $\omega_{QD}$ for
Sc$_2$O$_3$ and In$_2$O$_3$ (the sesquioxides with the smallest
lattice parameters) depart from the linear dependence on $a$ shown
by the rest of the sesquioxides. Concerning the asymmetry
parameter, while $\eta_D$ is nearly independent of $a$ for all the
measured sesquioxides and close to 0, $\eta_C$ does not show an
increase with $a$ for In- and Sc-sesquioxides, as occurs for the
rest of the systematics (see Fig.~\ref{fig:5}). The anomalous
behavior of the $\omega_{Q}$ and $\eta_C$ parameters may indicate
the presence of structural distortions in the host lattice induced
by the impurity. We will discuss this statement in the next
section.

\begin{figure}
\includegraphics*[bb= 46mm 166mm 114mm 267mm, viewport=-1.3cm -0.4cm 8.5cm 10cm,scale=0.9]{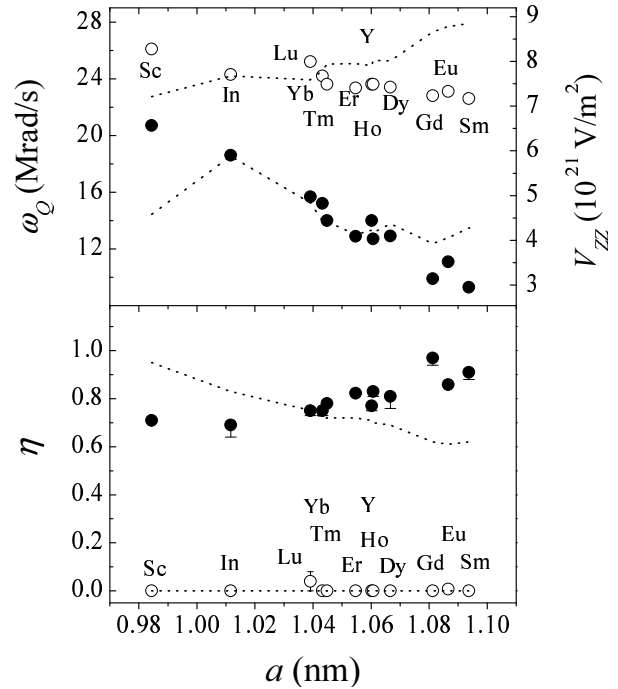}
\caption{\label{fig:6}  Experimental quadrupole frequency
$\omega_Q$ and asymmetry parameter $\eta$ for substitutional
$^{111}$Cd impurities at sites $C$ (black) and $D$ (hollow) in
bixbyites\cite{Errico99,Bibi84,Bartos91}$^-$\cite{Carbonari99}
plotted as a function of the lattice parameter $a$ at RT. The
results for Lu$_2$O$_3$ and Tm$_2$O$_3$ correspond to 1273 and 583
K, respectively (see Refs. \onlinecite{Errico99} and
\onlinecite{Carbonari99}). In order to obtain V$_{zz}$ from
$\omega_Q$ we used $Q_{Cd} = (+) 0.83(13)~b$ .\cite{Herzog80} The
errors are smaller than the symbols. Dotted lines are the PCM
predictions. The $\omega_Q$ predictions are normalized to the
In$_2$O$_3$ values.}
\end{figure}

If we compare the trend of $\omega_{Q}$ and $\eta$ with
point-charge model (PCM) predictions (see Fig.~\ref{fig:5}), we
find that the observed increase of $\eta_C$ and the dependence of
$\omega_{QC}$ and $\omega_{QD}$ on $a$ are in poor agreement. At
this point it is necessary  to analyze the experimental hyperfine
interaction parameters at $^{111}$Cd impurity sites in bixbyites
reported in the
literature.\cite{Errico99,Bibi84,Bartos91}$^-$\cite{Carbonari99}
Figure~\ref{fig:6} shows a monotonic dependence of the hyperfine
parameters $\omega_Q$ and $\eta$ on the lattice parameter $a$ for
both sites. These trends are in poor agreement with those
predicted by the PCM, as can be seen in the same figure. In the
past, this discrepancy was explained by two different approaches.
In the first one,\cite{Bartos93b,Lupascu94b} the PCM was
considered accurate enough to exactly predict the hyperfine
parameters that characterize the EFG  at Cd sites in bixbyites. In
this model, the EFG is given by

\begin{equation}\label{Vijionico}
V_{ZZ}^{ionic} = (1-\gamma_{\infty})V_{ZZ}^{latt},
\end{equation}
where $\gamma_{\infty}$ is the Sternheimer antishielding factor
and, in the principal axes system:

\begin{equation}\label{Vijlatt}
V_{ZZ}^{latt}= \frac {e}{4\pi\varepsilon_0} \sum_k Z_k   \frac
{3x_{ki} x_{kj} - \delta_{ij} r_k^2} {r_k^5}.
\end{equation}
In the above expression, $Z_k$ denotes the (arbitrary) ionic
charge, $x_{ki}$ and $x_{kj}$ the coordinates, and $r_k$ the
distance of the $k^{th}$ ion to the site where the EFG is
calculated, located at the origin of the coordinated system for
simplicity. The discrepancy between PCM predictions and the
experimental results was attributed to a wrong determination of
the atomic positions in these compounds (mainly the oxygen atoms).
Within this framework, the authors of Refs.
~\onlinecite{Lupascu94b} and ~\onlinecite{Bartos93b} used the PCM
to refine the atomic positions in order to reproduce the
experimental EFG results. It is worthy of mention that that
approach presented several assumptions, e.g., pure ionic bonding
between the probe atom and its neighbors was assumed and local
lattice distortions caused by the $^{111}$Cd impurity were not
considered. At this point, it is important to mention that that
model was proposed before a $^{181}$Ta systematics in bixbyites
was established. Now, the very different trend between the
experimental $^{181}$Ta data presented here and the PCM
predictions with the refined coordinates of
Ref.~\onlinecite{Bartos93b} gives clear evidence of the failure of
this model in the $^{181}$Ta case and makes its applicability
suspicious in the $^{111}$Cd one. In the next section, we present
theoretical evidence that supports this statement.

The second approach was first proposed to describe the whole set
of EFG results at $^{111}$Cd in binary oxides in a unified
way,\cite{Renteria92} and then it was extended to the $^{181}$Ta
impurity. Since this model was constructed to exactly reproduce
the magnitude of all the existent experimental EFG results, the
hypothesis involved in it and the predictions of this model for
the sign of $V_{zz}$ (opposite to those of $V_{ZZ}^{ionic}$, see
Ref. \onlinecite{Renteria92}) have to be discussed now,
enlightened by the first-principles predictions, which were not
available at the time the model was proposed. In this approach,
Renter\'{\i}a et al. \cite{RenteThesis,Renteria92} proposed the
existence of a dominating ¨local contribution¨ to the EFG. In this
model the EFG is given by

\begin{equation}\label{Vijsemiempirico} V_{ZZ} = V_{ZZ}^{local}
+ (1-\gamma_{\infty})V_{ZZ}^{latt}.
\end{equation}

The local contribution to the EFG takes into account covalence
between the probe and its neighbors. From the study of the
experimental EFG results at $^{111}$Cd and $^{181}$Ta impurity
sites in binary oxides we demonstrated that, for both probes,
$V_{ZZ}^{local}$ shows an almost linear dependence on
$V_{ZZ}^{ionic}$ (see Ref.~\onlinecite{Renteria99}). We also
demonstrated that $V_{ZZ}^{local}$ can be factorized in two parts,
one that depends on the geometry of the cationic site and another
that depends on the electronic configuration of the probe, thus
\cite{Renteria99,Errico2001}

\begin{eqnarray}
 V_{ZZ}(site~i, probe~i)&=& \mu^{probe~i}(1-\gamma_\infty)^{probe~i} \nonumber\\
 &\times& \mu^{geometry}~V_{ZZ}^{latt}(site~i).\label{Vzzexp3}
\end{eqnarray}

 We must say here that this
semiempirical model is still based on point-charge summations and
takes into account the contribution of the electronic structure of
the probe to the EFG by means of antishielding factors and
empirical parameters (such as $\mu^{probe}$ and $\mu^{geometry}$).
Moreover, this model does not take into account possible effects
introduced by the impurity character of the probe, such us lattice
distortions. Nevertheless, this simple semiempirical model has
been successfully used as a tool for the assignment of hyperfine
interactions in complex systems.\cite{Renteria2000} In the next
section the hypothesis and predictions of this model are discussed
in terms of very recent band-structure FP-LAPW calculations of the
EFG at Cd and Ta impurity sites in In$_2$O$_3$ and other binary
oxides.

\subsection{\label{sec:4.B}FP-LAPW results for the EFG tensor at impurity sites
in binary oxides}

Since this paper deals with the EFG at impurity sites, the
interpretation of the experimental results involves the
understanding of chemical differences between the probe atom and
the indigenous ion replaced by the impurity. The experimental
results show that the differences between probes and indigenous
atoms are manifest in subtle ways that are not well described by
conventional models, as already pointed out in Ref.
\onlinecite{Catchen94}. For an accurate calculation of the EFG,
the electronic configuration of the host, perturbed by the
presence of the impurity, has to be determined. This can be done
in the framework of the Density-Functional Theory (DFT). In this
kind of calculation, electronic and structural effects introduced
in the host by the presence of the impurity probe (impurity
levels, structural distortions, etc.) can be described without the
use of arbitrary suppositions. Unfortunately, these calculations
prove to be not trivial and time-consuming. For this reason, very
few calculations have been performed in systems with impurities
and the method is far from being routinely applied in this
field.\cite{PRL, PRB} Moreover, the applicability of the DFT to
bixbyites that contain rare earth cations is not clear yet since
the 4$f$-valence electrons are not well described by this
theory.\cite{Llois2004}

In the last years we performed FP-LAPW \cite{Wei85} calculations
in the systems TiO$_2$(Cd),\cite{PRL, PRB}
SnO$_2$(Cd),\cite{SnO22001} TiO$_2$(Ta), and
SnO$_2$(Ta).\cite{Pucon2003} We present here new calculations of
the EFG tensor at Cd and Ta impurities that replace cations in the
model case In$_2$O$_3$. For the calculations we employed the
Wien97.10 implementation\cite{Wien97} of the FP-LAPW method within
the Generalized Gradient Approximation (GGA) for the
exchange-correlation potential.\cite{GGA} For the parameter
RK$_{MAX}$, which controls the size of the basis set, we took the
value of 7 and introduced local orbitals for In-4$d$ and 4$p$,
O-2$s$, Cd-4$d$ and 4$p$, and Ta-6$s$, 5$p$, and 4$f$ orbitals.
With these parameters, the convergence errors in the $V_{ii}$
components are smaller than 0.1 $\times$ 10$^{21}$V/m$^2$. In the
case of the asymmetry parameter $\eta$, the convergence error is
smaller than 0.05. In order to study the structural distortions
introduced by the impurity in the host we considered the
displacements of the impurity nearest oxygen neighbors until
forces acting on these atoms vanished, assuming that structural
distortions preserve the point-group symmetry of the cell in its
initial configuration. Due to the large number of atoms in the
unit cell of In$_2$O$_3$, we did not use a super-cell approach for
the simulation of the isolated impurity. Details of the method of
calculation and the way to deal with the impurity were extensively
explained in Refs.~\onlinecite{PRL} and \onlinecite{PRB}. It is
clear that, even including the new FP-LAPW results reported here,
the number of EFG calculations at impurity sites is very small to
perform a systematic study, but some important conclusions can be
drawn.

The first important theoretical result is that in all the studied
systems the impurity probes introduce local distortions in the
host. It is worthy of mention that in all the cases the amount of
distortion per atom decreases rapidly from the impurity nearest
oxygen neighbors (O$_{NN}$) to other shells. In the case of Cd
located at both cationic sites of In$_2$O$_3$ we found that the
impurity produced relaxations of about 5$\%$ (see Table II) of the
cadmium O$_{NN}$. Additionally, in the case of Cd located at site
$C$ we found that the impurity moved out from the symmetry site,
but this displacement (3x10$^{-4}$ nm) is very small compared to
the displacement of the Cd O$_{NN}$. Larger relaxations were also
found in TiO$_2$ and SnO$_2$ for the Cd impurity. These
relaxations can be understood from the fact that the bond length
of the sixfold coordinated Cd ion in CdO is about 0.235 nm. It
seems that the local structure of the Cd impurity tries to
reconstruct the environment of Cd in its own oxide.\cite{SnO22001,
Pucon2003} Since in the rest of the bixbyites most of the bond
lengths are smaller than 0.235 nm, one could also expect
relaxations around the Cd impurity site in all of them. The
presence of these structural relaxations in bixbyites with Cd
shows that the refinement proposed in
Refs.~\onlinecite{Lupascu94b} and \onlinecite{Bartos93b} is not
valid since it is based on the fact that the Cd impurity does not
introduce structural distortions in the host. In this sense this
model mistakes a local effect (the O$_{NN}$ relaxations originated
by the impurity character of the probe) for a pretended wrong
determination of the host atomic positions by XRD, which involves
all the lattice.

In the case of Ta at site $D$ of In$_2$O$_3$, we found that Ta
induces a 7$\%$ contraction in the Ta-O$_{NN}$ distance.
Calculations at site $C$ predict 5-10$\%$ contractions of the
Ta-O$_{NN}$ distances (see Table II) and a negligible displacement
of Ta from the symmetry site. We have also found contractions in
the case of SnO$_2$(Ta), whereas in TiO$_2$(Ta) the local
structural distortions are negligible.\cite{Pucon2003} These
contractions can be understood from the fact that the bond lengths
of the sixfold coordinated Ta ion in TaO$_2$ are about 0.202 nm.
As in the case of Cd, it seems that the local structure tries to
reconstruct the environment of Ta in its oxide. Due to the larger
bond lengths of the rest of the bixbyites we should also expect
contractions of the Ta-O$_{NN}$ distances. From these
considerations it is clear that distortions with different
symmetry and/or magnitude for In- and Sc-sesquioxides with respect
to those of the rest of the systematics could be at the origin of
the anomalous behavior of the hyperfine parameters for Ta in
bixbyites shown in Fig.~\ref{fig:5}.

\begin{table*}
\caption{\label{table2}Experimental and FP-LAPW-refined In-O
distances (in nm) in pure In$_2$O$_3$ compared with the relaxed
Cd-O and contracted Ta-O distances (last two rows) calculated when
a Cd or a Ta atom replaces an In atom in In$_2$O$_3$.}
\begin{ruledtabular}
\begin{tabular}{lllll}
  & \multicolumn{1}{c}{Site $D$} & \multicolumn{3}{c}{Site $C$}\\
     & \multicolumn{1}{c}{$d(In-O)$} & \multicolumn{1}{c}{$d(In-O1)$} &
  \multicolumn{1}{c}{$d(In-O2)$} & \multicolumn{1}{c}{$d(In-O3)$}\\ \hline

 Pure In$_2$O$_3$ (Experimental)\cite{Marezio66}  & $0.219(2)$& $0.212(2)$ & $0.219(1)$ & $0.221(2)$ \\
 % \hline
Pure In$_2$O$_3$ (FP-LAPW, this work) & 0.218  & 0.213  & 0.219  & 0.223 \\

In$_2$O$_3$:Cd (FP-LAPW, this work)    & $0.227$ & $0.220$ &
$0.229$ & $0.233$ \\

In$_2$O$_3$:Ta (FP-LAPW, this work)    & $0.202 $ & $0.197$ &
$0.197 $ &
$0.209 $ \\

\end{tabular}
\end{ruledtabular}
\end{table*}

\begin{table*}
\caption{\label{table3} Experimental EFG results available for
TiO$_2$, SnO$_2$, and In$_2$O$_3$ doped with Cd and Ta compared
with the PCM and FP-LAPW predictions. The signs of $V_{ZZ}$ at the
Cd and Ta sites are not known experimentally.}
\begin{ruledtabular}
\begin{tabular}{lllcl}
\multicolumn{1}{c}{System} & \multicolumn{1}{c}{Method} &
\multicolumn{1}{c}{V$_{ZZ}$} & \multicolumn{1}{c}{V$_{ZZ}$
direction}& \multicolumn{1}{c}{$\eta$}  \\ \hline

TiO$_2:Cd$       & PCM              & $-2.21$    & [1$\overline{1}$0]           & $0.43$       \\
                 & FP-LAPW\cite{PRB}   & $+4.55$    & [110]                      & $0.26$       \\
                 & Experimental\cite{PRL}  & $5.34(1)$  & [110] or [1$\overline{1}$0]  & $0.18(1)$
                 \\\hline

SnO$_2:Cd$       & PCM                      & $+5.21$    & [110]       & $0.40$       \\
                 & FP-LAPW\cite{SnO22001}      & $+5.35$    & [110]       & $0.10$       \\
                 & Experimental\cite{Renteria91}   & $5.83(4)$  & -           & $0.18(2)$
                 \\\hline

In$_2$O$_3$:Cd (Site D)  & PCM                  & $+7.5$     & [111]  & $0.00$   \\
                         & FP-LAPW (this work)     & $+7.6$     & [111]  & $0.00$   \\
                         & Experimental\cite{Habe96}   & $7.7(1)$   & [111]\footnotemark[1]  & $0.00(5)$
                         \\\hline

In$_2$O$_3$:Cd (Site C)  & PCM                  & $-4.9$     & [0$\overline{0.75}$1]  & $0.83$  \\
                         & FP-LAPW (this work)     & $+5.6$     & [0$\overline{0.75}$1]  & $0.68$  \\
                         & Experimental\cite{Habe96}   & $5.9(1)$   & [0$\overline{1}$1]\footnotemark[1]     & $0.69(1)$    \\
                         \hline\hline

TiO$_2$:Ta       & PCM                          & $-4.55$    & [1$\overline{1}$0]  & $0.43$       \\
                 & FP-LAPW\cite{Pucon2003}         & $-13.0$    & [001]             & $0.69$       \\
                 & Experimental\cite{Renteria2004}     & $13.3(1)$  & [001]             & $0.56(1)$
                 \\\hline

SnO$_2$:Ta       & PCM                      & $+10.76$    & [110]  & $0.40$       \\
                 & FP-LAPW\cite{SnO22001}   & $-17.7$     & [001]  & $0.96$       \\
                 & Experimental\cite{Moreno91}     & $17.04(1)$  & -          & $0.70(2)$    \\\hline

In$_2$O$_3$:Ta (Site D)  & PCM                     & $+15.46$   & [111]    & $0.00$       \\
                         & FP-LAPW (this work)        & $+19.7$    & [111]         & $0.00$       \\
                         & Experimental\cite{Renteria97}  & $16.31(2)$ & -        & $0.149(7)$    \\\hline

In$_2$O$_3$:Ta (Site C)  & PCM                     & $-10.10$    & [0$\overline{0.75}$1]  & $0.83$       \\
                         & FP-LAPW (this work)        & $ -17.8 $      & $ [010.3] $                       & $0.26$       \\
                         & Experimental\cite{Renteria97}  & $12.35(9)$  & -                      & $0.60(9)$    \\

\end{tabular}

\end{ruledtabular}

\footnotetext[1]{~Experimental values obtained with $^{111}$Cd in
Er$_2$O$_3$ and Ho$_2$O$_3$ single crystals.\cite{Lupascu94b}}

\end{table*}

We can discuss now the FP-LAPW predictions for the EFG tensor and
compare them with the experiments, the PCM, and the semiempirical
model of Ref.~\onlinecite{Renteria99}. From our calculations we
found that relaxations (or contractions) beyond the O$_{NN}$ shell
do not produce qualitative changes in the EFG tensor. For this
reason all the FP-LAPW results for the EFG tensor shown here
correspond to the relaxed position of the O$_{NN}$. There is an
excellent agreement between the experiments and FP-LAPW
predictions for the EFG at Cd impurity sites, as shown in Table
III. On the other hand, the PCM fails completely in the
description of the EFG for TiO$_2$. The agreement is better for
SnO$_2$, but the PCM fails in the prediction of the asymmetry
parameter. This failure should be
  expected due to their covalency originated in the short Cd-O$_{NN}$ bond lengths.
  In the case of In$_2$O$_3$, where the bond lengths are larger, the PCM
agrees with the FP-LAPW EFG prediction for Cd at site $D$, but
disagrees in the sign of the major component of the EFG tensor for
Cd at site $C$.

In the case of the Ta impurity, the agreement between the
experiments and FP-LAPW predictions is very good, with exception
of Ta at site $C$ in In$_2$O$_3$. Here it is important to say that
due to the hypothesis used in the relaxation process (structural
relaxations preserve the point-group symmetry of the cell in its
initial configuration), FP-LAPW cannot reproduce the small
departure from the axial symmetry experimentally observed for Ta
at site $D$ of In$_2$O$_3$ (see Table III). On the other hand, the
PCM fails completely in the description of the EFG at Ta
impurities located at cationic sites of TiO$_2$ and SnO$_2$. In
the case of the EFG at Ta impurities located at sites $C$ and $D$
of In$_2$O$_3$, the PCM agreement with the available experimental
data (V$_{ZZ}$ and $\eta$) is better than that of the FP-LAPW
predictions. Hence, from eight probe-oxygen configurations, five
are not well described by the PCM.

The {\it ab initio} calculations predicted  local distortions in
almost all the cases studied. If we use the relaxed positions of
the O$_{NN}$ atoms predicted by FP-LAPW in the PCM calculations,
this model still fails in the predictions of the EFG. Moreover, in
most of the cases the disagreement becomes worse. From this
discussion it is clear now that the failure of the PCM prediction
of the EFG at impurity sites is not only due to the use of wrong
atomic positions but to a wrong description of the electronic
structure of the impurity-host system, in particular in the
vicinity of the impurity.

Based on the FP-LAPW calculations we can finally discuss the
validity of the hypothesis on which the semiempirical model is
based, i.e., the existence of a
 dominating local contribution to the EFG.
 In this model, the EFG can be written as

\begin{equation}\label{Vzzmcp}
V_{ZZ}= V_{ZZ}^{local} - \gamma_\infty V_{ZZ}^{latt (PCM)} +
V_{ZZ}^{latt (PCM)},
\end{equation}
meanwhile in the FP-LAPW framework we have

\begin{eqnarray}\label{Vzzlapw}
V_{ZZ}^{LAPW}= V_{ZZ}^{val} +  V_{ZZ}^{latt (LAPW)}=\nonumber \\ =
V_{ZZ}^{val, {\it p}} + V_{ZZ}^{val, {\it d}} + V_{ZZ}^{latt
(LAPW)},
\end{eqnarray}
where $V_{ZZ}^{val, {\it p}}(V_{ZZ}^{val, {\it d}})$ is the
valence contribution of Cd- or Ta-p(d) orbitals inside the
muffin-tin sphere of the impurity and $V_{ZZ}^{latt (LAPW)}$ is
the total contribution originated in the charge density outside
that sphere.

 In both cases (Eqs. 6 and 7) the lattice contributions are negligible.
Taking into account the $d$ orbitals involved  in the
$\gamma_{\infty}$ factor (4{\it d} for Cd, 5{\it d} for Ta), we
can correlate the term -$\gamma_{\infty}V_{ZZ}^{latt (PCM)}$ of
Eq. 6 with the {\it d} valence contribution to the EFG in FP-LAPW
calculations (Eq. 7). In this way, the {\it p} valence
contribution of Eq. 7 has to be correlated with $V_{ZZ}^{local}$
in Eq. 6. With this ``bridge'' linking both models, we can return
to the results obtained in the FP-LAPW calculations. In all the
studied systems (including In$_ 2$O$_ 3$) and for both probes we
found that the {\it p} contribution (5{\it p} in the case of Cd,
6{\it p} for Ta) dominates over the {\it d} contribution. Based on
these results, we can conclude that the {\it ab initio}
calculations of the EFG tensor at impurity sites support the
hypothesis of the existence of a local and dominating contribution
to the EFG on which the semiempirical model is based. This model
gives a better qualitative description of the EFG than the PCM
because it includes a contribution to the EFG that is not taken
into account in the PCM. But the semiempirical model is based on
the use of the $\gamma_{\infty}$ factor and point-charge
summations that do not take into account the structural
distortions introduced by the impurity. These shortcomings could
produce wrong predictions in our model, such as the wrong
prediction of the EFG sign in various of the compounds shown in
Table III. An extensive first-principles study of the EFG at
impurity sites in binary oxides is essential in order to remove
these shortcomings.

\section{\label{sec:5}CONCLUSIONS}

In order to compare complete sets of EFGs at $^{181}$Ta and
$^{111}$Cd sites in bixbyite oxides we determined the
electric-quadrupole hyperfine interactions at $^{181}$Ta
impurities in Eu$_2$O$_3$ and Ho$_2$O$_3$. Including previous
results, we could compare a set of twelve oxides having the same
crystalline structure. Our study shows that in all these oxides,
after the implantation and subsequent high-temperature annealing,
the probes $^{181}$Hf occupy two sites that have been identified
as the two nonequivalent cation sites of the bixbyite structure.
We confirm a jump in the values of $\omega_{QC}$ and $\omega_{QD}$
that takes place for $a <$ 1.0391 nm, that is for lattice
parameters smaller than that of Lu-sesquioxide. Concerning the
asymmetry parameter, while $\eta_D \approx$ 0.0 is nearly
independent of $a$ for all the measured sesquioxides, $\eta_C$ for
indium and scandium sesquioxides departs from the tendency that
rules its behavior for the rest of the systematics. These
anomalous experimental behaviors, in combination with our FP-LAPW
predictions for Ta in In$_2$O$_3$ (and in the other binary
oxides), suggest that Ta induces contractions of its O$_{NN}$ in
all the bixbyites, but with different symmetry and/or magnitude in
the case of In$_2$O$_3$ and Sc$_2$O$_3$.

The experimental EFG data obtained at $^{111}$Cd and $^{181}$Ta
impurity sites in binary oxides available at present clearly show
that a single ionic model cannot describe the EFG, suggesting the
existence of a dominating nonionic contribution as stated in the
semiempirical model. This assertion is supported by our
band-structure FP-LAPW calculations at Cd and Ta impurity sites in
binary oxides, which show that the dominating contribution to the
EFG is originated in Cd-5$p$ and Ta-6$p$ states, respectively.
These results can now explain the dependence of the local EFG,
proposed in the semiempirical model, on the electronic
configuration of the probes. Future FP-LAPW calculations,
currently in progress, at impurity sites in oxides and, in
particular, in bixbyites, will certainly go deeper into the
understanding of structural and electronic properties at impurity
sites in semiconducting oxides.

\begin{acknowledgments}
It is a pleasure to recognize our colleague Prof. Dr. Patricia
Massolo, who shared with us the motivations of this research. We
greatly appreciate the valuable suggestions of Dr. R. Vianden and
Dr. T. Butz  during the design of the detectors and the
acquisition of spare parts of the TDPAC spectrometer. This work
was partially supported by Agencia Nacional de Promoci\'on
Cient\'{\i}fica y Tecnol\'ogica (ANPCyT) under PICT98 03-03727,
Consejo Nacional de Investigaciones Cient\'{\i}ficas y T\'ecnicas
(CONICET) under PIP006/98, and Fundaci\'on Antorchas, Argentina,
and the Third World Academy of Sciences (TWAS), Italy, RGA 97-057.
The neutron irradiations performed by the GKSS reactor FRG-1,
Germany, are kindly acknowledged. M.R., L.A.E, and A.G.B. are
members of CONICET, Argentina.
\end{acknowledgments}

%\bibliography{paperon4}% Produces the bibliography via BibTeX.

\begin{thebibliography}{}

\bibitem{KauVianden}E. N. Kaufmann and R. J. Vianden, Rev. Mod. Phys. {\bf 51}, 161 (1979).
\bibitem{Lerf}A. Lerf and T. Butz, Hyperfine Interact. {\bf 36}, 275 (1987).
 \bibitem{Abeti}A. R. L\'opez-Garc\'{\i}a, Magn. Res. Rev. {\bf 15}, 119 (1990).
\bibitem{Lupascu94a} D. Lupascu, M. Uhrmacher, and K. P. Lieb, J. Phys.: Condens. Matter
{\bf 6}, 10445 (1994).
\bibitem{Catchen94}J. M. Adams and G. L. Catchen, Phys. Rev. B {\bf 50}, 1264
(1994).
\bibitem{Lany2000}S. Lany, P. Blaha, J. Hamann, V. Ostheimer, H. Wolf, and
T. Wichert, Phys. Rev. B {\bf 62}, R2259 (2000); T. Wichert and S.
Lany, Hyperfine Interact. {\bf 136/137}, 453 (2001).
\bibitem{Evenson2000}N. Mommer, T. Lee, J. A. Gardner, and W. E.
Evenson, Phys. Rev. B {\bf 61}, 162 (2000).
\bibitem{Schatz86}T. Klas, J. Voigt, W. Keppner, R. Wesche, and G. Schatz,
Phys. Rev. Lett. {\bf 57}, 1068 (1986); R. Fink, T. Koch, G.
Krausch, J. Marien, A. Plewnia, B. U. Runge, G. Schatz, A. Siber,
and P. Ziemann, Phys. Rev. B {\bf 47}, R10048 (1993).
\bibitem{Ramallo2003}%J. M. Ramallo L\'opez, F. G. Requejo, A.
%G. Bibiloni, M. Renter\'{\i}a, L. Gutierrez, and E. E. Mir\'o, Z.
%Naturforsch. A {\bf 55}, 327 (2000);
J. M. Ramallo-L\'opez, M. Renter\'{\i}a, E. E. Mir\'o, F. G.
Requejo, and A. Traverse, Phys. Rev. Lett. {\bf 91}, 108304
(2003).
\bibitem{Pasque83}A. F. Pasquevich, F. H. S\'anchez, A. G. Bibiloni, J.
Desimoni, and A. L\'opez-Garc\'{\i}a, Phys. Rev. B {\bf 27}, 963
(1983); J. Desimoni, A. G. Bibiloni, L. Mendoza-Z\'elis, A. F.
Pasquevich, F. H. S\'anchez, and A. L\'opez-Garc\'{\i}a, Phys.
Rev. B {\bf 28}, 5739 (1983).
\bibitem{Errico99}L. A. Errico, M. Renter\'{\i}a, A. G. Bibiloni, and F. G. Requejo,
Hyperfine Interact. {\bf 120/121}, 457 (1999).
\bibitem{Bibi84}A. G. Bibiloni, J. Desimoni, C. P. Massolo, L. Mendoza-Z\'elis, A. F. Pasquevich, F. H. S\'anchez, and A.
L\'opez-Garc\'{\i}a, Phys. Rev. B {\bf 29}, R1109 (1984).
\bibitem{Habe96}S. Habenicht, D. Lupascu, M. Uhrmacher, L.
Ziegeler, K. P. Lieb, and ISOLDE Collaboration, Z. Phys. B {\bf
101}, 187 (1996).
\bibitem{Achtziger93}N. Achtziger and W. Witthuhn, Phys. Rev. B
{\bf 47}, 6990 (1993).
\bibitem{Vianden1988}R. Vianden and U. Feuser, Phys. Rev. Lett. {\bf 61}, 1981
(1988).
\bibitem{Renteria2000}M. Renter\'{\i}a , L. A. Errico, A. G. Bibiloni,
K. Freitag, and F. G. Requejo, Z. Naturforsch. A {\bf 55}, 155
(2000).
\bibitem{Forker2003}M. Forker, S. Muller, P. de la Presa, and A. F. Pasquevich
Phys. Rev. B {\bf 68}, 14409 (2003).
\bibitem{Rots2001}J. Meersschaut, C. Labbe, M. Rots, and S. D.
Bader, Phys. Rev. Lett. {\bf 87}, 107201 (2001).
\bibitem{Lerf1997}A. Lerf and T. Butz, Angew. Chem. Int. Ed. Engl. {\bf 26}, 110
(1987).
\bibitem{Pasque81}  A. F. Pasquevich, A. G. Bibiloni, C.
P. Massolo, F. H. S\'anchez, and A. L\'opez-Garc\'{\i}a, Phys.
Lett. A {\bf 82}, 34 (1981).
\bibitem{Wiarda92} See, for example, D. Wiarda,
M. Uhrmacher, A. Bartos, and K. P. Lieb, J. Phys.: Condens. Matter
{\bf 5}, 4111 (1993); M. Neubauer, A. Bartos, K. P. Lieb, D.
Lupascu, M. Uhrmacher, and Th. Wenzel, Europhys. Lett. {\bf 29},
175 (1995); J. Luthin, K. P. Lieb, M. Neubauer, M. Uhrmacher, and
B. Lindgren, Phys. Rev. B {\bf 57}, 15272 (1998), and references
therein.
\bibitem{Bolse88} W. Bolse, A. Bartos, J. Kesten, M. Uhrmacher,
and K. P. Lieb, XXIII Zacopane School on Physics, edited by K.
Krolas and K. Tomala (Institute of Nuclear Physics, Cracow, 1988).
\bibitem{Kesten90} J. Kesten, W. Bolse, K. P. Lieb, and M. Uhrmacher, Hyperfine
Interact. {\bf 60}, 683 (1990).
\bibitem{RenteThesis}  M. Renter\'{\i}a, Ph.D. thesis, Universidad
Nacional de La Plata, Argentina, 1992.
\bibitem{Renteria92} M. Renter\'{\i}a,
C. P. Massolo, and A. G. Bibiloni, Mod. Phys. Lett. B {\bf 6},
1819 (1992).
\bibitem{Weht94} R. Weht, G. Fabricius, M. Weissmann, M. Renter\'{\i}a, C.
P. Massolo, and A. G. Bibiloni, Phys. Rev. B   {\bf 49}, 14939
(1994).
\bibitem{PRL}L. A. Errico, G. Fabricius, M. Renter\'{\i}a, P. de la Presa, and M. Forker,
 Phys. Rev. Lett. {\bf 89}, 55503 (2002).
\bibitem{PRB}L. A. Errico, G. Fabricius, and  M.
 Renter\'{\i}a, Phys. Rev. B {\bf 67}, 144104 (2003).
\bibitem{Shitu98}J. Shitu, A.F. Pasquevich, A.G. Bibiloni, M. Renter\'{\i}a, F.G.
Requejo, Mod. Phys. Lett. B {\bf 12}, 281 (1998).
\bibitem{Moreno91}  M. S. Moreno, J. Desimoni, A. G. Bibiloni, M. Renter\'{\i}a,
 C. P. Massolo, and K. Freitag, Phys. Rev. B {\bf 43}, 10086 (1991).
\bibitem{Bartos91} A. Bartos, K. P. Lieb,
A. F. Pasquevich, M. Uhrmacher, and ISOLDE collaboration, Phys.
Lett. A {\bf 157}, 513 (1991).
\bibitem{Shitu92} J. Shitu, D. Wiarda, Th. Wenzel, M. Uhrmacher,
K. P. Lieb, S. Bedi, and A. Bartos, Phys. Rev. B {\bf
46}, 7987 (1992).
\bibitem{Lupascu94b} D. Lupascu, A. Bartos, K. P. Lieb, and M.
Uhrmacher, Z. Phys. B {\bf 93}, 441 (1994).
\bibitem{Pasque} A. F. Pasquevich, A. M.
Rodr\'{\i}guez, H. Saitovich, and P. R. de Jesus-Silva, private
communication.
\bibitem{Carbonari99}A. W. Carbonari,
J. Mestnik-Filho, R. N. Attili, M. Moralles, and R. N. Saxena,
Hyperfine Interact. {\bf 120/121}, 475 (1999).
\bibitem{Pasque94} A. F. Pasquevich, A. G. Bibiloni, C. P.
Massolo, M. Renter\'{\i}a, J. A. Vercesi, and K. Freitag, Phys.
Rev. B {\bf 49}, 14331 (1994).
\bibitem{Renteria97}M. Renter\'{\i}a, F. G. Requejo, A. G. Bibiloni, A. F. Pasquevich, J. Shitu, and K. Freitag, Phys.
Rev. B {\bf 55}, 14200 (1997).
\bibitem{Renteria98} M. Renter\'{\i}a, A. G. Bibiloni, F. G. Requejo,
A. F. Pasquevich, J. Shitu, L. A. Errico, and K. Freitag, Mod.
Phys. Lett. B {\bf 12}, 819 (1998).
\bibitem{Renteria99} M. Renter\'{\i}a, K. Freitag, and L. A. Errico,
Hyperfine Interact. {\bf 120/121}, 449 (1999).
 \bibitem{Errico2001} L. A. Errico, M.
Renter\'{\i}a, A. F. Pasquevich, A. G. Bibiloni, and K. Freitag,
Eur. Phys. J. B {\bf 22}, 149 (2001).
\bibitem{Germantulio}G. N. Darriba, L. A. Errico, and M Renter\'{\i}a, in {\it Anales AFA 14},
proceedings of the 87$^{th}$ Reuni\'on Nacional de F\'{\i}sica,
Huerta Grande, 2001, p. 208, edited by Asociaci\'on F\'{\i}sica
Argentina (Universidad Nacional del Centro,Tandil, 2002).
\bibitem{Eyring79} L. Eyring, in {\it Handbook on the Physics and Chemistry of
Rare Earths}, edited by K. A. Gschneidner and L. Eyring
(North-Holland, Amsterdam, 1979), p. 337.
\bibitem{Frauen} H. Frauenfelder and R. Steffen,
in {\it $\alpha$-, $\beta$-, and $\gamma$-Ray Spectroscopy},
edited by K. Siegbahn (North-Holland, Amsterdam, 1968), Vol. 2, p.
917; G. Schatz and A. Weidinger, in {\it Nuclear Condensed Matter
Physics. Nuclear Methods and Applications}, translated by J. A.
Gardner (John Wiley $\&$ Sons, Chichester, 1996), p. 63.
\bibitem{Mendoza77} L. A. Mendoza-Z\'elis, A. G.
Bibiloni, M. C. Caracoche, A. R. L\'opez-Garc\'{\i}a, J. A.
Mart\'{\i}nez, R. C. Mercader, and A. F. Pasquevich, Hyperfine
Interact. {\bf 3}, 315 (1977).
\bibitem{Butz83} T. Butz and A. Lerf, Phys.
Lett. A {\bf 97}, 217 (1983).
\bibitem{DRXN} Crystal structure data used in PCM predictions. Two
references are quoted if lattice parameters and atomic positions
come from different articles (lattice parameter articles are
quoted first). Sc$_2$O$_3$: R. W. G. Wyckoff, {\it Crystal
Structures}, (Wiley Interscience, New York, 1964), Vol. 2;  R.
Norrestam, Ark. Kemi. {\bf 29}, 343 (1968); In$_2$O$_3$: M.
Marezio, Acta Crystallogr. {\bf 20}, 723 (1966); Y$_2$O$_3$,
Dy$_2$O$_3$, and Ho$_2$O$_3$: E. N. Maslen, V. A. Streltsov, and
N. Ishizawa, Acta Crystallogr. B {\bf 52}, 414 (1996);
Er$_2$O$_3$: see Ref. ~\onlinecite{Lupascu94b}; Yb$_2$O$_3$: T.
Schleid and G. Meyer, J. Less-Comm. Met. {\bf 149}, 73 (1989);
Tm$_2$O$_3$: H. Ishibashi, K. Shimomoto, and K. Nakahigashi, J.
Phys. Chem. Solids {\bf 9}, 809 (1994); Lu$_2$O$_3$, Eu$_2$O$_3$,
Gd$_2$O$_3$, and Sm$_2$O$_3$: S. Stecura and W. Campbell, Thermal
Expansion and Phase Inversion of Rare Earth Oxides, Bureau of
Mines Report of Investigations N$^o$ 5847, U.S. Department of the
Interior, 1961; A. Saiki, N. Ishizawa, N. Mizutani, and M. Kato,
Yogyo Kyokai Shi {\bf 93}, 649 (1985).
\bibitem{Herzog80} P. Herzog, K. Freitag, M. Reuschenbach, and H. Walitzki, Z.
Phys. A {\bf 294}, 13 (1980).
\bibitem{Bartos93b} A. Bartos, K. P. Lieb, M. Uhrmacher, and
D. Wiarda, Acta Crystallogr. B {\bf 49}, 165 (1993).
\bibitem{Llois2004} V. Vildosola, A. M. Llois, and J. G. Sereni, Phys. Rev. B
{\bf 69}, 125116 (2004).
\bibitem{Wei85} S. H. Wei and H. Krakauer, Phys. Rev. Lett. {\bf 55}, 1200 (1985).
\bibitem{SnO22001} L. A. Errico, G. Fabricius, and  M.
 Renter\'{\i}a, Hyp. Interact. {\bf 136/137}, 749 (2001).
\bibitem{Pucon2003}L. A. Errico, G. Fabricius, and M. Renter\'{\i}a,
Phys. Stat. Sol. (b) 241, 2394 (2004).
\bibitem{Wien97}P. Blaha, K. Schwarz, P. Dufek, and J. Luitz, {\sc wien97},
Vienna University of Technology, 1997. Improved and updated Unix
version of the original copyrighted {\sc Wien}-code, which was
published by P. Blaha, K. Schwarz, P.I. Sorantin, and S. B.
Trickey, in Comput. Phys. Commun. {\bf 59}, 399 (1990).
\bibitem{GGA} J. P. Perdew, K. Burke, and M. Ernzerhof, Phys. Rev.
Lett. {\bf 77}, 3865 (1996).
\bibitem{Marezio66} M. Marezio, Acta Crystallogr. {\bf 20}, 723
(1966).
\bibitem{Renteria91} M.
Renter\'{\i}a, A. G. Bibiloni, M. S. Moreno, J. Desimoni, R. C.
Mercader, A. Bartos, M. Uhrmacher, and K. P. Lieb, J. Phys.:
Condens. Matter {\bf 3}, 3625 (1991).
\bibitem{Renteria2004}M. Renter\'{\i}a, G. N. Darriba, L. A. Errico, and
P. D. Eversheim, to be published.
\end{thebibliography}

\end{document}